\documentstyle[12pt,epsf]{article}

\def\be{\begin{equation}}
\def\ee{\end{equation}}
\def\bea{\begin{eqnarray}}
\def\eea{\end{eqnarray}}
\font\extra=msbm10 scaled \magstep1
\def\bbb #1{\hbox{{\extra #1}}}
\begin{document}
\thispagestyle{empty}
\baselineskip=20pt
\

\begin{center}
{\LARGE{\bf The finite difference algorithm}}
\smallskip

{\LARGE{\bf for higher order supersymmetry}}
\vskip1cm

B. Mielnik${}^{1,2}$, L.M.~Nieto${}^3$, and 
O. Rosas--Ortiz${}^{1,3}$
\vskip0.5cm

{\it
${}^1$Departamento de F\'{\i}sica, CINVESTAV-IPN, AP
14-740
\\
07000 M\'exico DF, Mexico
\\  [2ex]
${}^2$Institute of Theoretical Physics, 
Warsaw University, Ho\.{z}a 69 
Warsaw, Poland
\\  [2ex]
${}^3$Departamento de F\'{\i}sica Te\'orica, Universidad de
Valladolid 
\\
47011 Valladolid, Spain
}
\end{center}
\vskip1cm

\begin{abstract}
\baselineskip=20pt
The higher order supersymmetric partners of the 
Schr\"odinger's Hamiltonians can be explicitly
constructed by iterating a  simple finite difference
equation corresponding to the B\"{a}cklund transformation.
The method can completely replace the   Crum determinants.
Its  limiting, differential case   offers some new
operational advantages. 
\end{abstract}

\vglue 1.5cm 

\begin{description}
\item[PACS:]  03.65.Ge, 03.65.Fd, 03.65.Ca
\item[Key-Words:] Supersymmetry, factorization, difference 
equations
\end{description}

\vglue 1cm 

{\bf Accepted for publication in Phys. Lett. A (2000)}

\newpage
\baselineskip=15pt

\section{Introduction}

It is notable that while the exact solutions in  
general relativity fill already ample handbooks, the 
exactly  solvable problems of quantum theory form still an
exceptional area. Yet, it seems  that they are just the
``top of an iceberg'', an opinion recently confirmed by
advances in the exact methods  
\cite{Zak97}. One of the simplest ways of obtaining new 
exactly solvable spectral problems in Schr\"odinger's
quantum mechanics  is to look for pairs of quantum
Hamiltonians coupled ``supersymmetrically'' by an {\it
intertwining operator\/} $A_1$ 
\be  
H_1 A_1 = A_1 H_0.
\label{I.1}
\ee 
This equation should hold on a certain dense domain of the 
Hilbert space  of states $\cal H$, assuring that the
spectrum of
$H_0$ yields some information about the spectrum of $H_1$ 
and {\it vice versa\/}.  The best known cases of (\ref{I.1})
occur in  
${\cal H} = { L}^2({\bbb R})$, where $H_0$ and $H_1$ are
the Schr\"odinger Hamiltonians
\be 
H_0 = -\frac{1}{2}\frac{d^2}{dx^2} + V_0(x),
\qquad H_1 = -\frac{1}{2}\frac{d^2}{dx^2} + V_1(x),
\label{I.2}
\ee 
and $A_1$, $A_1^{\dagger}$ are the first order differential 
operators 
\be 
A_1= \frac{d}{dx} + \beta_1(x), \qquad A_1^{\dag} = 
-\frac{d}{dx}  +
\beta_1(x).
\label{I.3}
\ee
By introducing (\ref{I.2}) and (\ref{I.3}) into 
(\ref{I.1}), and after an elementary integration, one gets
the Riccati equation for $\beta_1(x)$
\be 
-\beta'_1(x) + \beta_1^2(x) = 2[V_0(x) - \epsilon],
\label{I.4}
\ee
where $\epsilon$ is an arbitrary integration constant 
called the  {\it factorization energy\/} (a terminology
which does not imply that $\epsilon$ must be a  physical 
energy).  Simultaneously, 
\be  
V_1(x) = V_0(x) + \beta'_1(x).
\label{I.5}
\ee
Equations (\ref{I.4})  and (\ref{I.5}) are necessary and 
sufficient conditions for the Hamiltonians to be factorized
as
\be 
H_0 - \epsilon = \frac12\, A_1^{\dag} A_1,  \qquad H_1 - 
\epsilon =
\frac12\, A_1 A_1^{\dag}.
\label{I.6}
\ee
The susy partners (\ref{I.5})--(\ref{I.6}) appear in all 
quantum problems solvable by factorization \cite{classics}. 
An element of arbitrariness in this construction 
was explored in 1984 \cite{Mie84}; its presence in
Witten's susy quantum mechanics was noticed by Nieto 
\cite{Nie84}; links with the historical Darboux theory
\cite{Dar882} were established by Andrianov {\it et al\/}
\cite{And84}. 

The higher order intertwining was considered by Crum 
\cite{Cru55} (see also Krein \cite{Kre57}, Adler and Moser
\cite{Moser}). In physical applications though, for a long 
time, the attention was focused mostly on the first order
intertwining operator $A_1$ (see \cite{Alv88} and references
quoted therein). The idea of the higher order $A_1$
reemerged in 80'tieth
\cite{FerT84,Suk85} and it was systematically pursued since 
1993 (see the works of Veselov and Shabat \cite{shabat2},
Andrianov {\it et al\/} \cite{And95}, Eleonsky {\it et al\/}
\cite{Dub92}, Bagrov and Samsonov \cite{Bag95}, Fern\'andez
{\it et al\/}
\cite{Fer97,FGN}, and one of us \cite{Ros98}). As it turns 
out, the higher order susy partners of an arbitrary $H_0$
can be generated just by iterating the Darboux
transformation with different factorization constants
$\epsilon$ (see interesting theorems by Bagrov and Samsonov
in \cite{Bag95}). Formally, this can be done by introducing
a higher order intertwiner constructed by means of the {\it
determinant formula\/} of Crum \cite{Cru55}, a method
fundamentally important, though not very easy to apply. The
purpose of this note is to explore a faster method based on
the B\"acklund transformation \cite{Rog84} which operates
very simply at the level  of the $\beta$-functions ({\it
superpotentials\/}) in equations (\ref{I.3})--(\ref{I.5}).
It consist just in applying the finite difference calculus,
without caring about wave functions,  differential
equations, etc.  Moreover, if applied to a sufficiently rich
initial family of solutions to (\ref{I.4}), it permits to
recover the complete parametric dependence in all subsequent
susy steps without the need of integrating  any differential
equations 
\cite{Zak90,Mat91,Sta95}. We also show that the 
differential  case of the algorithm gives an entire susy 
hierarchy for a fixed $\epsilon$ just in terms of 
quadratures and  algebraic operations.   

\vglue 1cm

\section{The finite difference algorithm}

The solutions of the Riccati equation
(\ref{I.4}), and the potentials (\ref{I.5}), depend
implicitly on the factorization energy $\epsilon \in  {\bbb
R}$.  We shall make explicit this dependence by writing
$\beta_1(x,\epsilon)$ for the solutions to (\ref{I.4}), and
$V_1(x,\epsilon)$ for the corresponding susy partners of
$V_0(x)$. Hence, the intertwiner in (\ref{I.1})  reads
$A_1(\epsilon)= d/dx + \beta_1 (x, \epsilon)$, while the 
corresponding Hamiltonians are $H_0=(1/2) A_1^{\dagger}
(\epsilon) A_1(\epsilon)  +
\epsilon$, and $H_1(\epsilon)=(1/2) A_1(\epsilon)  
A_1^{\dagger} (\epsilon) + \epsilon$. Notice that $H_0$ does
not depend on 
$\epsilon$, yet it admits many different factorizations for
many factorization energies $\epsilon$, each one defining 
a different intertwined Hamiltonian $H_1(\epsilon)$,
with a different potential $V_1(x, \epsilon)$.

Let us fix now $\epsilon = \epsilon_1$, and consider one of 
these  potentials:
\[
V_1(x,\epsilon_1) = V_0(x) + \beta_1'(x,\epsilon_1).  
\]
Since $V_1(x,\epsilon_1)$ is a susy partner of $V_0(x)$,  
the eigenfunctions $\psi_1$ of the Hamiltonian
$H_1(\epsilon_1)$ can be generated from the eigenfunctions
$\psi_0$ of $H_0$ by  using the operator $A_1(\epsilon_1)$.
It means simply that if
$\psi_0(x,\epsilon)$ is a sufficiently smooth function and 
if
$\psi_1(x,\epsilon) \propto A_1(\epsilon_1)\psi_0 (x,
\epsilon)$, then:
\be 
H_0 \psi_0(x,\epsilon) = \epsilon\psi_0(x,\epsilon)  
\Longrightarrow H_1(\epsilon_1) 
\psi_1(x,\epsilon)= \epsilon\psi_1(x,\epsilon) . 
\label{fd2} 
\ee
Notice that (\ref{fd2}) does not require the square 
integrability of either
$\psi_0$ or $\psi_1$.  Nevertheless, if $\epsilon$ 
corresponds  to one of the discrete spectrum eigenvalues $E$
of $H_0$, then
$\psi_0$ can be chosen square integrable.

A question arises now, can $V_1(x,\epsilon_1)$ be a 
convenient  point of  departure for the next susy step? The
answer is positive and admits an  explicit construction. In
geometric form it was described by Adler  as a mapping of
complex polygons corresponding to the B\"acklund 
transformations \cite{adle} (see also Veselov and Shabat
\cite{shabat2}).  Its presence in the  supersymmetric
algorithm can be shown directly,  using only the Riccati
equation (\ref{I.4}). Indeed,  if $\beta_1(x,\epsilon)$ is a
solution of (\ref{I.4}), the  function 
\[
u_1(x,\epsilon) =
\exp{\left[-\int\beta_1(x,\epsilon) dx\right]}
\] 
satisfies $H_0 u_1(x,\epsilon) =
\epsilon \, u_1(x,\epsilon) $. Therefore, the function 
defined as
\be
u_2(x,\epsilon_1, \epsilon) = A_1(\epsilon_1) 
u_1(x,\epsilon)  = [\beta_1(x,\epsilon_1) -
\beta_1(x,\epsilon)]\exp \left[ -\int
\beta_1(x,\epsilon)dx \right]
\label{udos}
\ee
satisfies $H_1(\epsilon_1) u_2(x, \epsilon_1, \epsilon)= 
\epsilon\, u_2(x,
\epsilon_1,
\epsilon) $, and the function $\beta_2(x, \epsilon_1,
\epsilon)$ defined by 
\be
u_2(x, \epsilon_1, \epsilon) = \exp \left[-\int \beta_2(x, 
\epsilon_1,
\epsilon) dx \right]
\label{fd3}
\ee
must fulfill the new Riccati equation: 
\be
-\beta_2'(x, \epsilon_1, \epsilon) + \beta_2^2(x, 
\epsilon_1, 
\epsilon) = 2[V_1(x,\epsilon_1) - \epsilon]. 
\label{rdos}
\ee
Reading back (\ref{fd3}) and using (\ref{udos}) one has 
$$
\beta_2(x, \epsilon_1, \epsilon) = -\frac{d}{dx}\ln{u_2(x, 
\epsilon_1,
\epsilon)} = \beta_1(x,\epsilon) - 
\frac{\beta_1'(x,\epsilon_1) -
\beta_1'(x,\epsilon)}{\beta_1(x,\epsilon_1)  - 
\beta_1(x,\epsilon)},
$$ 
and using again (\ref{I.4}) one ends up with an algebraic 
expression
\be
\beta_2(x, \epsilon_1, \epsilon) = -\beta_1(x,\epsilon_1) -
\frac{2(\epsilon_1 - \epsilon)}{\beta_1(x,\epsilon_1) -
\beta_1(x,\epsilon)},
\label{fd4}
\ee 
which is the basic element of the auto-B\"acklund 
transformation  for the Korteweg-deVries (KdV) equation (as
reported in
\cite{Wahlq}, see also
\cite{Rog84} and \cite{Lam80}). Thus, the $\beta$-functions
of the second 
susy step are algebraically determined by a {\it finite
difference\/} operation performed on the $\beta$-functions 
of  the previous step.  They give the next operator
$A_2(\epsilon_1,
\epsilon)  \equiv d/dx + \beta_2(x,\epsilon_1,\epsilon)$ 
intertwining the Hamiltonian $H_1(\epsilon_1)$ with a new 
one
$H_2(\epsilon_1, \epsilon)
\equiv -(1/2) \, d^2/dx^2 + V_2(x, \epsilon_1, \epsilon)$,
\be
H_2(\epsilon_1, \epsilon) \, A_2(\epsilon_1, \epsilon)= 
A_2(\epsilon_1,
\epsilon) \, H_1(\epsilon_1),
\label{idos}
\ee
where
\be
V_2(x,\epsilon_1, \epsilon) = V_1(x,\epsilon_1) + 
\beta_2'(x, 
\epsilon_1,
\epsilon).
\label{2susy}
\ee

Quite obviously (\ref{I.1}) and (\ref{idos})  imply 
\[
H_2(\epsilon_1, \epsilon) \, {\cal
A}_2 = {\cal
A}_2 \, H_0,
\]
where ${\cal A}_2 = A_2(\epsilon_1, \epsilon) \, 
A_1(\epsilon_1)$.  In
\cite{shabat2,adle} the method has been applied to deform 
the  cyclic supersymmetric chains, but we shall show that it
can replace completely all other techniques in constructing
quite arbitrary susy sequences. 

Indeed, the method can be now repeated by induction,  
leading to the sequence of  first order intertwining
operators 
\be 
A_k (\epsilon) \equiv \frac{d}{dx} + \beta_k(x,\epsilon), 
\qquad k=1,2,...,n,
\label{fd6}
\ee 
where each $\beta_k,\, k\geq2,$  is simply the result of a 
finite difference  operation performed on $\beta_{k-1}$:
\be
\beta_k (x,\epsilon) = - \beta_{k-1} (x,\epsilon_{k-1}) - 
\frac{2(
\epsilon_{k-1} - \epsilon)}{\beta_{k-1} (x,\epsilon_{k-1}) - 
\beta_{k-1} (x,\epsilon)}. 
\label{fd7}
\ee
We have adopted here a shortcut notation making explicit 
only   the dependence of $A_k$ and $\beta_k$ on the
factorization constant introduced in the very last step,
keeping implicit the dependence on the previous
factorization constants (henceforth, the same criterion will
be used for any other symbol depending on
$k$ factorization energies):
\[
A_k(\epsilon) \equiv A_k(\epsilon_1, ...,\epsilon_{k-1},
\epsilon) ,\quad \beta_k(x,\epsilon) \equiv
\beta_k(x,\epsilon_1, ...,\epsilon_{k-1}, \epsilon).
\]
The $\beta_k$ functions constructed at each next susy step 
automaticaly solve the Riccati equation with the potential
$V_{k-1}$ of the previous step 
\be
-\beta_k'(x, \epsilon) + \beta_k^2(x, \epsilon) =
2[V_{k-1}(x,\epsilon_{k-1}) - \epsilon],
\label{fd8}
\ee
or equivalently 
\be
-\beta_k'(x,\epsilon) + \beta_k^2(x,\epsilon) = 
\beta_{k-1}'(x,\epsilon_{k-1}) + 
\beta_{k-1}^2(x,\epsilon_{k-1})  +  2(\epsilon_{k-1} -
\epsilon)
\label{A.0}
\ee
permitting to define the new potential 
\be
V_{k}(x,\epsilon) = V_{k-1}(x,\epsilon_{k-1}) + \beta_{k}' 
(x, 
\epsilon).
\label{fd9}
\ee
In the resulting sequence of the new Hamiltonians
\be
H_k (\epsilon) = -\frac12 \frac{d^2}{dx^2} + 
V_{k}(x,\epsilon) =
\frac{1}{2} A_k(\epsilon) A^{\dagger}_k(\epsilon) + 
\epsilon, 
\qquad k=1,2,...,n
\label{HK}
\ee
each next is intertwined with the previous one,
\be
H_k  (\epsilon) \, A_k (\epsilon)  = A_k (\epsilon)  \,
H_{k-1} (\epsilon_{k-1}),\qquad H_0  (\epsilon_{0})  \equiv
H_0.
\label{fd12}
\ee
All of them are the higher order susy 
partners of the initial $H_0$
\be 
H_k (\epsilon) \, {\cal A}_k = {\cal A}_k \, H_0,
\label{0K}
\ee
(compare with \cite{Moser}) where ${\cal A}_k$ is the 
higher  order intertwining operator obtained in the
factorized form:
\be
{\cal A}_k \equiv A_k (\epsilon)  A_{k-1} (\epsilon_{k-1}) 
\cdots A_2 (\epsilon_2)  A_1 (\epsilon_1).
\label{tocho1}
\ee
(Here we follow our notation, making explicit the 
dependence of  each $A_k$ on its last factorization energy.)
Once having the sequence of  superpotentials $\beta_k (x,
\epsilon)$, constructed by the algorithm  (\ref{fd7}), it is
easy to determine  the new eigenfunctions $u_k(x,\epsilon)$
injected by the process  into  the Hamiltonians 
$H_{k-1}(\epsilon_{k-1})$. Indeed, the $u_k(x, \epsilon_k)$
are  defined by the first order differential equations
\be
A_k (\epsilon) \, u_k(x, \epsilon) = \frac{d}{dx} u_k(x, 
\epsilon) +
\beta_k(x, \epsilon) u_k(x, \epsilon) = 0,  \qquad
k=1,2,...,n. 
\label{fd15}
\ee
Up to now we have maintained implicit the element of 
arbitrariness  in the $\beta$-functions at every  recurrence
step. However, the true  advantages of the method can be
seen if
$\beta_1(x, \epsilon)$ is wider  than just a particular 
solution of (\ref{I.4}). Indeed, for each fixed 
$\epsilon$, we have to our disposal the general solution
of the Riccati  equation (\ref{fd8}), {\it i.e.\/}, an 
entire one-parametric family of  functions. Thus, {\it
e.g.\/}, the general solution of (\ref{I.4}),  for a given
$\epsilon_1$,  depends on an additional integration
parameter $\lambda_1$.  Given any particular solution
$\beta_{1p} (x,\epsilon_1)$, the general one is obtained by
an elementary transformation: 
\be
\beta_1(x,\epsilon_1)  = \beta_{1p} (x, \epsilon_1)-
\frac{d}{dx} 
\ln
\left[ \lambda_1- \int \exp\left\{ 2\int^x \beta_{1p}(y, 
\epsilon_1) \, dy\right\} dx
\right] \equiv \beta(x, \epsilon_1, \lambda_1).
\label{fd16}
\ee 
Now, the algorithm (\ref{fd4})  
can be interpreted as an operation on two {\it classes of
functions\/} $\beta_1(x,\epsilon)$ and 
$\beta_1(x,\epsilon_1)$:  an arbitrary member of the class
$\beta_1(x,\epsilon)$ subtracted from an arbitrary member
of $\beta_1(x,\epsilon_1)$ in the denominator of
(\ref{fd4}). Quite similarly, the general iteration step
(\ref{fd7}) can be applied for
$\beta_{k-1}(\epsilon)$ and $\beta_{k-1}(\epsilon_{k-1})$ 
meaning any two superpotentials of the previous step, the
only condition stating that they must solve the
corresponding Riccati equation with the same potential
$V_{k-2}$ (a fact which is automaticaly assured in the
iterative process). If so, the method accumulates the
integration constants, permitting to recover the full
parametric dependence  on the $n$-th susy potentials in a
purely algebraic way. Thus,  {\it e.g.},
$\beta_2(x,\epsilon) :=
\beta(x; \epsilon_1,
\epsilon;  \lambda_1, \lambda)$ with the dependence on 
$\epsilon_1,
\lambda_1, \epsilon, \lambda$ inherited from the previous 
step ($\lambda_1$ and $\lambda$ selecting two solutions
among the  elements of the families $\beta_1(x,\epsilon_1)$
and
$\beta_1(x,\epsilon)$ respectively).  In general, our 
algorithm yields $\beta_k(x,\epsilon) :=
\beta(x; \epsilon_1,..., \epsilon_{k-1}, \epsilon;  
\lambda_1,...,
\lambda_{k-1},\lambda)$ where the parameters $\epsilon,  
\lambda$ will be varied to perform the next step, whereas
$\epsilon_1,...,
\epsilon_{k-1},
\lambda_1,..., \lambda_{k-1}$ will stay fixed. In what 
follows,  we shall employ the symbols $\beta_k(x,\epsilon)$
whenever wanting to abbreviate, but we shall hold the
dependence on the accumulated integration constants
$\epsilon_1,..., \epsilon_{k-1}, \lambda_1,...,
\lambda_{k-1}$ if necessary. Let us now introduce the
functions: 
\be
\Omega_k (x, \epsilon) =
\cases{ \beta_1(x, \epsilon), \quad 
\mbox{\rm for $k=1$};\cr  \cr
\beta_k (x,\epsilon) + \beta_{k-1} (x, \epsilon_{k-1}), 
\quad 
\mbox{\rm for $k \geq 2$}. \cr}
\ee
Observe their  
{\it fractal\/} structure accumulating the 
{\it finite difference\/} derivatives (compare with 
\cite{shabat2})
\be
\Omega_k(x,\epsilon_k) = -\frac{ 2 (\epsilon_{k-1} - 
\epsilon_k) }{
\Omega_{k-1} (x, \epsilon_{k-1}) - \Omega_{k-1}(x, 
\epsilon_k) }, 
\quad
\quad k\geq 2.
\label{fd17}
\ee
Now, (\ref{fd7}) can be rewritten as
$\beta_k (x,\epsilon) = \Omega_k (x, \epsilon) - 
\beta_{k-1} (x,
\epsilon_{k-1} )$. For example, for $k=2$ we get 
(\ref{fd4}),  while for
$k=3$ we have
\begin{eqnarray}
\hskip-0.9cm \beta(x; \epsilon_1, \epsilon_2, \epsilon_3; 
\lambda_1,
\lambda_2,
\lambda_3) 
\!\!&\!\!= \!\!&\!\!
\beta(x;\epsilon_1; \lambda_1) +
\frac{2(\epsilon_{1}-\epsilon_{2})}{
\beta(x;\epsilon_1;\lambda_{1})- \beta(x;\epsilon_2; 
\lambda_{2})}
\nonumber
\\    [1ex]
\!\!&\!\!  \!\!& \!\!\! +
\frac{\epsilon_{2}-\epsilon_{3}}{ \displaystyle
\frac{\epsilon_{1}-\epsilon_{2}}{
\beta(x;\epsilon_1;\lambda_{1})-\beta(x;\epsilon_2;
\lambda_{2})}  -
\frac{\epsilon_{1}-\epsilon_{3}}{ \displaystyle
\beta(x;\epsilon_1;\lambda_{1})- \beta(x;\epsilon_3;
\lambda_{3})}}.
\label{fractal}
\end{eqnarray}
As one can also notice, the ``$n$-susy potentials'' 
$V_n(x)$,  present the different fractal structures for $n$
even and odd.  Summarizing:  
\be
\label{fd18} 
V_n(x,\epsilon_n) = V_0 (x)  + \sum_{k=1}^{n} \beta_k'(x,
\epsilon_k) 
= V_0 (x)  + \sum_{k=1}^{n} \left[\cos^2 
\left(\frac{(k+n)\pi}{2}\right)
\right] \frac{d}{dx} \Omega_k (x,\epsilon_k). 
\ee
This expression has been already applied to simplify the 
study  of $n$-susy partners of the harmonic oscillator
\cite{Fer97} and the Coulomb potential
\cite{Ros98}. Notice now that when $\lambda_k =
\lambda_{k-1}$, and taking the limit $\epsilon_{k-1} \to 
\epsilon$ in (\ref{fd7}), the method produces a differential
algorithm (a {\it confluent susy\/} operation) 
\be
\beta^{\rm conf}_{k}(x,\epsilon) = - 
\beta_{k-1}(x,\epsilon) - {2}\left(  \frac {\partial
\beta_{k-1}(x,\epsilon)}{\partial 
\epsilon}
\right)^{-1} ,        
\label{2.15}
\ee 
which can be freely iterated. Likewise, the finite 
differences  can be replaced by the $\epsilon$-derivatives
in one or more places of the fractal formulae
(\ref{fractal}). The mechanism has been implicitly applied
by Stalhoffen (see Section III of
\cite{Sta95}), in the context of Crum's determinant. 

\section{Simple and confluent operations}

The difference-differential algorithms (\ref{fd7}), 
(\ref{2.15})   simplify remarkably 
the construction of transparent wells 
\cite{Zak97,Mat91,Sta95}.  In fact, the Riccati 
equation (\ref{I.4}) for $V_0(x)=0$ has 
the general solution 
\be
\beta_1(x;\epsilon;\alpha)= -\sqrt{2\epsilon}\, 
\cot[\sqrt{2\epsilon}\,  (x-\alpha)],  
\label{free1}
\ee
with four different real 1-susy branches: S  ({\it
singular}), R  ({\it regular}),  
P ({\it periodic}), and N ({\it null}), for different  (real
or  complex)  values of $\epsilon$ and 
$\alpha$ (see Table~I). For example, the regular case (R) 
leads  to the modified P\"oschl-Teller potential 
$V_1^{R}(x,\epsilon,b)= -\kappa^2/\cosh^2[\kappa(x+b)]$, 
while  the null case (N) corresponds to the potential 
barrier
$V_1^{N}(x, \epsilon=0,a)  = (x-a)^{-2}$. All of them are
particular cases of the Weierstrass
$\wp$-function: 
\be
\label{pfunction}
V_1(x;\epsilon;\alpha)=
\wp(x+\alpha,3(4\epsilon/3)^2,(4\epsilon/3)^3)+
\frac{2\epsilon}{3}.
\ee

\vskip0.3cm

\begin{center}

\begin{minipage}{12cm}
{\footnotesize
TABLE I. The four different real superpotentials $\beta_1$ 
coming out from (\ref{free1}), depending on the values of
$\epsilon$ and the integration parameter $\alpha$. In each 
case S means singular, R regular,
${\rm P}$ periodic, and N null. The parameters $a$ and
$b$ are arbitrary real numbers.}
\end{minipage}
\medskip

\begin{tabular}{ccccc}
\cline{1-5} \\ [-2ex]
\cline{1-5} 
& & & &   \\ [-1ex]
\quad Case \qquad & $\epsilon$ & $\sqrt{2\epsilon}$ &
\quad $\alpha $  & $\beta_1(x,\epsilon)$  \\ [1ex] 
\hline
& & & &   \\ [-1ex]
S & \quad$\epsilon<0$ \quad &\quad $i\sqrt{2|\epsilon|}= 
i\kappa$ 
\quad& $a$ & \quad$-\kappa\, \coth[\kappa(x-a)]$  
\quad \\ [2ex]
R & $\epsilon<0$ & $i\sqrt{2|\epsilon|}=i\kappa$ & \quad
$\displaystyle -b-\frac{i\pi}{2\kappa}$ & $-\kappa\,
\tanh[\kappa(x+b)]$\quad \\ [2ex]
${\rm P}$ & $\epsilon>0$ & $\sqrt{2\epsilon}= k$ & $a$ & 
$-k\, \cot[k(x-a)]$      \\   [2ex]
N & $0$ & $0$ & $a$  &  $\displaystyle -\frac1{x-a}$ \\ [2ex]
\cline{1-5}  \\ [-2ex]
\cline{1-5} 
\end{tabular}
\end{center}
\vskip0.5cm

By applying the 
algorithm (\ref{fd4}) to the superpotentials R and S as
given in Table~I, one immediately gets the transparent 
double wells in form of the 
{\it quadratic Bargmann potentials} \cite{Lam80} 
(see Figure 1)  
\be
\label{free2}
V_2 (x,\epsilon_2) = - (\kappa_1^2 - \kappa_2^2) \, \, 
\frac{
\kappa_1^2 \, \, {\rm csch}^2 \, [ \kappa_1 (x +b)] +   
\kappa_2^2
\, \, {\rm sech}^2 \, [ \kappa_2 (x -a) ] }{ (\,
 -\kappa_1 \coth
\, [ 
\kappa_1 (x +b)  ] + \kappa_2 \tanh \, [ \kappa_2 (x -a) ] 
\,)^2 }. 
\ee
\begin{figure}[htbp]
\centerline{
\epsfxsize=10cm
\epsfbox{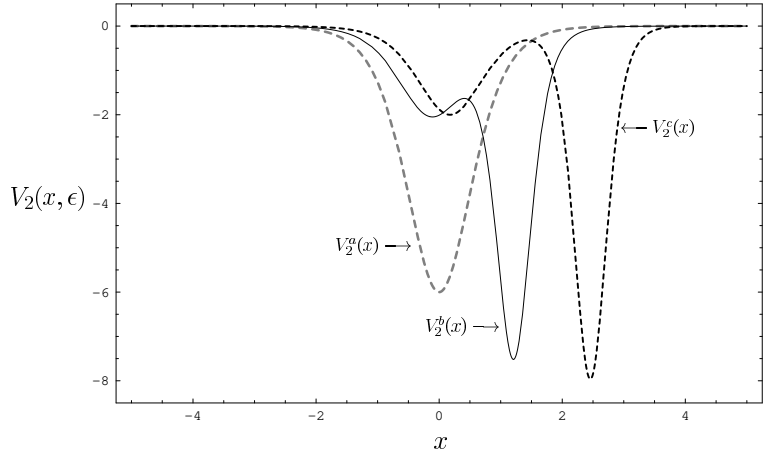}}
\begin{center}
\begin{minipage}{12cm}
\caption{\footnotesize
The double transparent wells (\ref{free2})  illustrate an application of
the auto-B\"acklund transformation (\ref{fd4}). The figure shows different
members of the 2-susy family $V_2 (x, \epsilon_2, \epsilon_1)$ for
$\epsilon_1 = -4$, $\epsilon_2=-1$, with integration parameters (a)
$a=b=0$, (b) $a=0.254$, $b=-1.018$, and (c)  $a=0.565$, $b=-2.262$. 
}
\end{minipage}
\end{center}
\end{figure}

The double wells (\ref{free2}) are well 
known in supersymmetry, but the use of the B\"acklund 
transformation  makes the derivation considerably simpler,  
explaining also their place  
in the theory of the KdV equation,  
$v_t -6v v_x + v_{xxx}=0$.  
As a matter of fact, all the transparent $n$-susy 
wells can be now reinterpreted as the {\it instantaneous 
forms}  of  the multisoliton solutions propagating according
to KdV   (thus, {\it e.g.\/}, the 2-susy potentials
(\ref{free2})  originate   the propagating soliton pairs
\cite{Cooper}).  Presumably, the susy  
partners of an arbitrary $V_0 (x)$ generate as well  the 
traveling ``solitonic deformations" on a more general, 
non-null   background.              

The operational advantages of the differential 
algorithm (\ref{2.15}) deserve as much attention. The 
application of   the finite difference algorithm to the  
null wells is so easy,   because for $V_0=0$ one knows the
solutions of the Riccati  eq. (\ref{I.4})  for any
$\epsilon$. For non-null 
$V_0$,  (\ref{fd4}) can be applied efficiently only if 
one knows   at least two solutions of (\ref{I.4}) for two
different  ``factorization energies" $\epsilon$ (compare
\cite{Fer97}, 
\cite{Ros98}). For a general 
$V_0$, however, this is not granted. It is therefore 
interesting, that the difficulty is circumvented by the 
differential  algorithm (\ref{2.15}) which gives an easy 
access to higher order  susy steps even if one starts from
just one superpotential 
$\beta_k(x,\epsilon)$ 
for a fixed $\epsilon$. Indeed, by taking the limit 
$\epsilon \rightarrow \epsilon_{k-1}$, one reduces 
(\ref{A.0})  to the following equation for 
$\beta_k = \beta_k(x,\epsilon=\epsilon_{k-1})$: 
\be
-\beta_k' + \beta_k^2 = \beta_{k-1}' + \beta_{k-1}^2,  
\quad  k=2,3,...,
\label{A.1}
\ee
which has the particular solution $\beta_k = -\beta_{k-1}$ 
and  integrates easily yielding: 
\begin{eqnarray}
\beta_k \!&\! =\! &\! - \beta_{k-1} + 
          \frac{{ e}^{-2\int \beta_{k-1}dx}}
        {\Gamma_k - \int { e}^{-2 \int\beta_{k-1} dx} dx}
            \nonumber \\
        \!&\! =\! &\! -\beta_{k-1} - \frac{d}{dx}
            \ln \left[ \Gamma_k  - 
                   \int {{ e}^{-2\int \beta_{k-1}dx}dx} 
\right] .
\label{A.2}
\end{eqnarray}

Note that the right hand side of (\ref{A.2}) depends 
essentially   just on one integration constant. The
recurrence (\ref{A.1})  acquires  even simpler form in  
terms of the new `key functions' 
$B_n=\exp(-\int \beta_n dx + c_n)$ (the integration 
constants 
$K_n=\exp c_n$ are superfluous to determine $\beta_n$).  
In fact, integrating both sides of (\ref{A.2}), taking to 
the   exponent, differentiating again, and choosing  
properly the  (inessential) constants $K_n$, one has: 
\be
     \frac{d}{dx}(B_nB_{n-1})= - B_{n-1}^2 .
\label{A.3}
\ee
When this method is applied to 
$V_0(x)=x^2/2$ with $\epsilon =-1/2$, the first 
step recovers the Abraham-Moses oscillator \cite{Mie84}, 
while the  second step leads to the  new 2-parametric 
family 
\be
V_2(x)=\frac{x^2}2-\frac{d^2}{dx^2} \ln \left[ 
\Gamma_2-\int_0^x dy\ e^{y^2}\left( \Gamma_1-
\frac{\sqrt{\pi}}2\ {\rm erf\,}x
\right)^2
\right],
\label{oscilconf}
\ee
which are different from the potentials discussed in
\cite{FGN}, because they appear for a fixed $\epsilon$ and 
do not add 2 new  spectral levels to $V_0$; we therefore
propose to call them the  {\it 2-nd order Abraham-Moses
potentials}.  The detailed analysis shows  
that (\ref{oscilconf}) are singular unless 
$\Gamma_1=\frac12 \,
\sqrt{\pi}$ and $\Gamma_2\geq \frac14 \,\sqrt{\pi}\,  \ln
2$, or 
$\Gamma_1=-\frac12 \,
\sqrt{\pi}$ and $\Gamma_2\leq -\frac14 \,\sqrt{\pi}\,  \ln
2$. A nonsingular case  with $\Gamma_1=\frac12 \,
\sqrt{\pi}$ and $\Gamma_2= 0.308$ is
shown in Figure~2. The higher order Abraham-Moses functions
can be as easily  obtained in terms of  
(\ref{A.2}). Generically, our algorithm 
is close to the  nice idea of Leble
\cite{Leble} (though Leble did not construct the  
non-trivial  contributions for
$\delta=\epsilon-\epsilon_{k-1}=0$).  
\begin{figure}[htbp]
\centerline{
\epsfxsize=10cm
\epsfbox{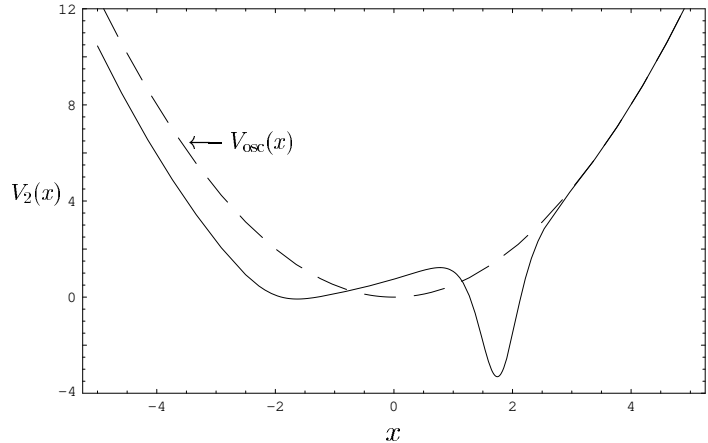}}
\begin{center}
\begin{minipage}{12cm}
\caption{\footnotesize
The solid line is a nonsingular 2-susy confluent partner for the harmonic
oscillator with factorization energy $\epsilon=-1/2$ (equation
(\ref{oscilconf})), for $\Gamma_1= \sqrt{\pi}/2$ and $\Gamma_2= 0.308$.
The dashed line represents the harmonic oscillator potential $x^2/2$. 
}
\end{minipage}
\end{center}
\end{figure}

An equally interesting application of the confluent formula 
(\ref{2.15}) arises for  
the periodic superpotential ${\rm P}$ in Table~I. In this
case, the first order susy partner is given by
\be
\label{1conf}
V_1^{\rm conf}(x, \epsilon) = V_1(x, \epsilon)  =
\frac{k^2}{\sin^2[k(x-a)]},
\ee
while the corresponding second order susy step leads to 
\be
\label{free3}
V_2^{\rm conf}(x, \epsilon) = 8 k^2 \, \, \frac{ 1 -\cos\,  
[2k (x-a)] - k (x-a) \, \sin \, [2k (x-a)] } {(\, \sin \, [2
k(x-a)] -2 k (x-a) \, )^2 }. 
\ee
Notice that (\ref{1conf}) is a periodic potential
with singularities at the points $x_n= a + 2n/k$, 
$n\in\bbb Z$,  while 
$V_2^{\rm conf}(x, \epsilon)$ has only one
singularity. The {\it confluent susy step\/} 
(\ref{2.15}) has removed all the singularities from  
$V_1^{\rm conf}(x,
\epsilon)$ except that placed on $x=a$. The potential 
(\ref{free3})  has been plotted on Figure~3 (dashed curve)
for
$a=7, k=1/\sqrt2$. Observe that we have rederived the 
Stahlhofen function, but the method is simpler (see 
equations (50) and (40) of \cite{Sta95}). Moreover, some
other potentials  with  analogous properties can be
immediately obtained.
Our Figure~3 reports the fourth order confluent
potential $V_4^{\rm conf}(x,
\epsilon)$, with $a=-7, k=1/\sqrt2$.  Notice
the appearance of two new symmetric singularities.

\begin{figure}[htbp]
\centerline{
\epsfxsize=10cm
\epsfbox{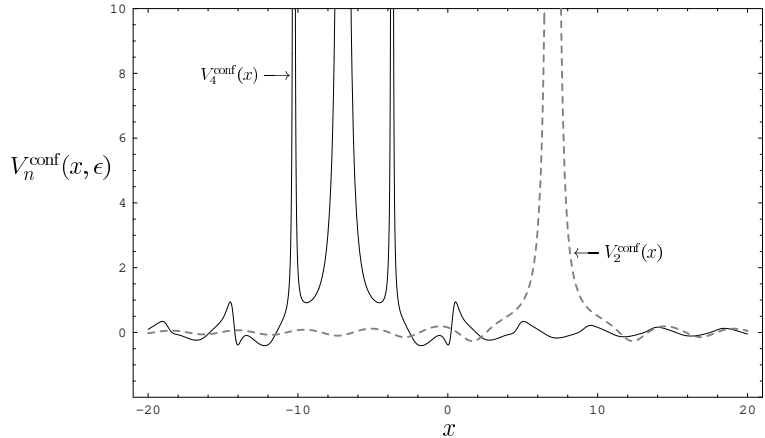}}
\begin{center}
\begin{minipage}{12cm}
\caption{\footnotesize
Examples of null, confluent $n$-susy potentials descending from the
(P)-branch:  $V_2^{\rm conf}(x)  \equiv V_2^{\rm conf}(x; 0.25; 7)$, and
$V_4^{\rm conf}(x)  \equiv V_4^{\rm conf}(x; 0.25; -7)$. For $n$ even in
(\ref{fd18}), the number of singularities grows with $n$ producing
potentials of Stalhoffen type. The odd cases have always an infinite
number of singularities. 
}
\end{minipage}
\end{center}
\end{figure}

As becomes obvious, for the periodic superpotentials  
$\beta_1$ in Table~I, the method leads to two distinct
classes of $n$-th order confluent potentials. For $n$ even,
the periodic superpotential $\beta_1$ does not contribute
directly to the sum (\ref{fd18}) (the term with $k=1$
vanishes). The resulting susy partners have only a finite
number of singularities. On the other hand, for $n$ odd, the
function $\beta_1$ contributes to the first term in the sum
(\ref{fd18}) and its global effect is never canceled. The
corresponding susy partners $V_1^{\rm conf}(x,\epsilon)$
therefore have infinite sequences of singularities.  This
difference reflects the distinct fractal structure of 
(\ref{fd18}) for $n$ even and odd. The possibility of mixed
applications of (\ref{fd7}) and (\ref{2.15})  is open.

\section*{Acknowledgements} 

This work has been  supported by the Spanish DGES
(PB94-1115 and PB98-0370) and by Junta de Castilla y 
Le\'on (CO2/199). BM and ORO acknowledge the kind
hospitality at Departamento  de F\'{\i}sica Te\'orica, Univ.
de Valladolid.  
ORO acknowledges support from CONACyT (Mexico). 
LMN thanks Prof. M.L. Glasser for  useful comments.


{\footnotesize{

}}

\newpage

\end{document}